\documentclass[%
 reprint,    
 amsmath,amssymb,
 aps,
]{revtex4-2}

\usepackage[pdftex]{graphicx}
\usepackage{dcolumn}
\usepackage{bm}

\usepackage{textcomp}

\def\dd{{\rm d}}

\def\jj{{\rm j}}
\def\ee{{\rm e}}
\def\degree{\mbox{$^\circ$}}
\def\sub#1{_{\rm #1}}

\def\mum{\,{\textmu m}}
\def\degree{\textdegree}

\begin{document}


\title{Extension of Babinet's relations to reflective metasurfaces:
Application to the simultaneous control of wavefront and polarization}

\author{Takayoshi Fujikawa}
\affiliation{Department of Electronic Science and Engineering, Kyoto University, Kyoto 615-8510, Japan.}
\author{Toshihiro Nakanishi}%
\affiliation{Department of Electronic Science and Engineering, Kyoto University, Kyoto 615-8510, Japan.}
\email{t-naka@kuee.kyoto-u.ac.jp}

\date{\today}

\begin{abstract}

We extend Babinet's relations, which are originally derived for transmissive metasurfaces,
to reflective metasurfaces embedding planar metallic structures 
in a substrate containing a reflection mirror. 
To verify the extended Babinet's relations, 
we investigate two types of reflective metasurfaces embedding self-complementary structures,
which produce $\pi$-phase difference between two orthogonal linear polarizations 
under specific conditions required for the extended Babinet's relations.
Both theoretical and experimental studies 
demonstrate a metasurface-based half-wave plate with a reflective metasurface 
including single self-complementary structures in the terahertz regions. 
In addition, we study a reflective metasurface with embedded self-complementary structures
with phase gradient in reflection,
and demonstrate simultaneous implementation of anomalous reflection and polarization conversion
with high efficiency.

\end{abstract}

\maketitle

 
\section{Introduction\label{sec:intro}}  

Metamaterials and metasurfaces are artificial media composed of subwavelength structures known as meta-atoms and
can be utilized to control electromagnetic wave propagation and polarization\cite{Yu2014,Miroshnichenko2015,Walia2015}.
The characteristics of metamaterials depend crucially on the structural symmetry 
of the meta-atoms \cite{Padilla_2007}. 
For example, metamaterials with broken mirror-symmetry structures
exhibit chiral responses \cite{Gonokami_2005,Rogacheva_2006,Wang_2009,Caloz_2020_1,Caloz_2020_2},
while those with anisotropic structures exhibit birefringence.
Various types of anisotropic metamaterials have been proposed 
to realize the functions of quarter and half-wave plates,
which induce a phase difference between the two orthogonal linear polarizations
\cite{Pors_2011,Pors_2013,Jiang_2014,Pfeiffer2014,Kruk_2016,Deng_2022,Nouman2016,Wang2015}.
The wavefront of the incident waves can be controlled in addition to the polarization
by introducing a phase gradient into the metasurface \cite{Yu2011,Ding2018}.
Simultaneous control of the wavefront and polarization has been previously reported 
using spatial distribution of various anisotropic meta-atoms with different orientations
\cite{Lin_2014,Yang_2014,Ding_2015,Arbabi_2015,Zheng_2015,Khorasaninejad_2016}.

Single-layer transmissive metasurfaces composed of planar metallic structures 
have been extensively investigated to control transmission waves in various frequency regions,
and Babinet's principle has been widely employed in the design of planar structures \cite{Falcone2004, Nakata_2013}.
The transmission coefficients of the original metallic structures for linear polarization
and its complementary structures for orthogonal polarization are linked 
by simple equations, referred to as Babinet's relations. 
Babinet's relations were applied to the design of reconfigurable polarization devices
using metasurfaces incorporated with phase changing materials, 
which are introduced to switch between metallic structures and their complementary structures
\cite{Nakata2016,Nakata2019,Nakanishi2020,Urade2022}.
In addition, unique properties of self-complementary structures can be applied 
to frequency-independent responses\cite{Urade2015} and quarter-wave plates \cite{Baena2015}.
Single-layer transmissive metasurfaces, however,
suffer substantial reflection loss, as discussed in Sec.~\ref{sec:babinet}.
On the other hand,
reflective metasurfaces composed of planar metallic structures and a reflective mirror
provide a solution to significantly improve the efficiency.
In this paper, we compare a transmissive metasurface composed of planar metallic structures
with a reflective metasurface composed of the same metallic structures embedded in a substrate
and a reflection mirror,
and derive conditions in which relations similar to conventional Babinet's relations are satisfied
for reflective metasurfaces in the latter part of Sec.~\ref{sec:babinet}.
  
According to the extended Babinet's relations,
the difference between the reflection phases for two orthogonal linear polarizations is always $\pi$
for reflective metasurfaces with self-complementary structures under the derived conditions.
In other words, the reflective metasurface functions as a half-wave plate. 
To test this hypothesis,
we design a reflective metasurface composed of self-complementary structures and 
a reflective mirror, which operates in the terahertz regime, 
and demonstrate a metasurface-based half-wave plate both experimentally and 
using electromagnetic simulations in Sec.~\ref{sec:single}.
Furthermore, we apply the extended Babinet's relations to phase-gradient metasurfaces
with spatially varying reflection phases.
We design a reflective metasurface that includes eight types of self-complementary structures,
thus realizing a linear phase gradient in a specific direction,
and demonstrate an anomalous reflection with polarization conversion caused by phase retardation
between the two orthogonal polarizations in Sec.~\ref{sec:gradient}.
This simultaneous control of wavefront  and polarization is a unique property derived from
the extended Babinet's relations.
The extended Babinet's relations may pave the way for efficient multifunctional 
metasurfaces such as reconfigurable intelligent surfaces \cite{Yang2022,elight_Li,PIER_Liu,PIER_Jin}
and polarization-encoded metasurfaces \cite{Lee2019,Ding2021}.


\section{Extension of Babinet's relations to reflective metasurface\label{sec:babinet}}

\begin{figure}[h] 
  \begin{center}
    \includegraphics{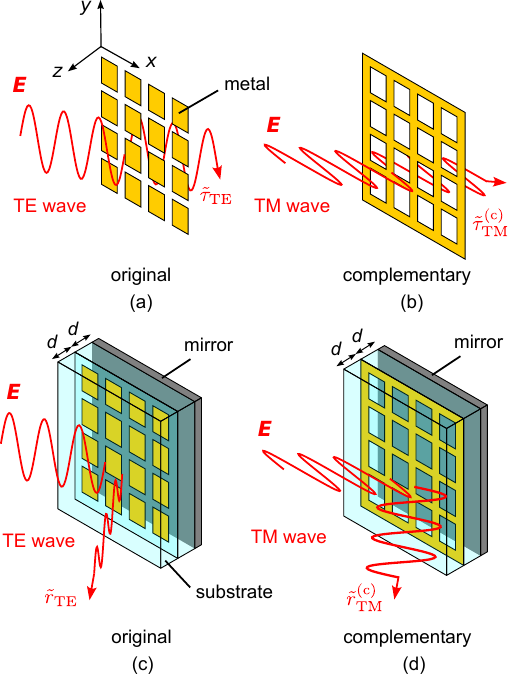}
    \caption{Electromagnetic response of (a) planar metallic structures and (b) complementary structures. 
    Electromagnetic response of a reflective metasurface embedding (c) planar metallic structures and (d) complementary structures.}
    \label{fig:babinet-meta}
  \end{center}     
\end{figure}

First, the conventional Babinet's relations for transmissive metasurfaces are reviewed.
Suppose a metasurface with thin metallic structures located on the $z=0$ 
plane is illuminated by TE waves with $y$-polarization, as shown in Fig.~\ref{fig:babinet-meta}(a).
Additionally, we assume that the anisotropic axes are aligned in $x$- and $y$-directions
and the periodicity of the structures is sufficiently small such that diffraction does not occur.
Under these conditions, the incident TE waves
are separated into two paths as transmitted TE waves
and reflected TE waves without diffraction.
Next, we consider the interaction between TM waves and a complementary metasurface
obtained using the impedance inversion given by $Z\sub{s}^{\rm (c)}=Z^2/[4 Z\sub{s}(x, y)]$,
where $Z$ is the wave impedance of the surrounding medium and
$Z\sub{s}(x, y)$ denotes the local impedance of the original metasurface shown 
in Fig.~\ref{fig:babinet-meta}(a).
If the original metasurface comprises a perfect electric conductor $Z\sub{s}(x, y)=0$
and the insulation holes $Z\sub{s}(x, y) \rightarrow \infty$,
the complementary metasurface can be obtained by swapping the metallic and insulating parts,
as shown in Fig.~\ref{fig:babinet-meta}(b).
It is known that 
Babinet's principle links the transmission coefficient $\tilde{\tau}\sub{TE}$ for the original metasurface
with  the transmission coefficient $\tilde{\tau}\sub{TM}^{\rm (c)}$ for the 
complementary metasurface as follows \cite{Nakata_2013}:
\begin{align}
  \tilde{\tau}\sub{TE} + \tilde{\tau}\sub{TM}^{\rm (c)} = 1. \label{Babinet_trans}
\end{align}
A similar relationship is derived by exchanging the TE and TM waves as  
\begin{align}
  \tilde{\tau}\sub{TM} + \tilde{\tau}\sub{TE}^{\rm (c)} = 1. \label{Babinet_trans2}
\end{align}
These relations are referred to as Babinet's relations for transmissive metasurfaces.

In this paper, we propose the reflective metasurface shown in Fig.~\ref{fig:babinet-meta}(c).
The metasurface embeds metallic planar structures identical to
Fig.~\ref{fig:babinet-meta}(a) in a substrate with the refractive index $n$ whose backside is covered with a metallic mirror.
The reflection coefficient $\tilde{r}_{i}$ ($i={\rm TE, TM}$) in  Fig.~\ref{fig:babinet-meta}(c)
is connected to the transmission coefficient $\tilde{\tau}_{i}$ and reflection coefficient $\tilde{\rho}_i$
of the metallic planar structure as shown in Fig.~\ref{fig:babinet-meta}(a).
The same relationship holds for complementary structures, as shown in Fig.~\ref{fig:babinet-meta}(b)
and a reflective metasurface embedding the structure, as shown in Fig.~\ref{fig:babinet-meta}(d).
In this section, first, the normal incidence of waves is considered for simplicity.
From analyses using transfer matrices, which are presented in Appendix \ref{sec:normal},
we obtain the reflection coefficient $\tilde{r}_{i}$ under a condition $kd=\pi/2$,
where $k$ is the wavenumber in the substrate, as follows:
\begin{align}
    \tilde{r}_i = - \frac{(n+2)\tilde{\tau}_i-2}{(n-2)\tilde{\tau}_i+2},
    \label{ri}
\end{align}
which shows that 
the reflection coefficient for Fig.~\ref{fig:babinet-meta}(c) is expressed as a M\"{o}bius transformation
of the transmission coefficient $\tilde{\tau}_i$ for the embedded metallic structures.
Under the condition $n=2$, the above relationship is reduced to 
\begin{align}
  \tilde{r}_i=-2 \left(\tilde{\tau}_i -\frac{1}{2}\right). \label{map}
\end{align}
The same relationship is expected for a reflective metasurface shown in Fig.~\ref{fig:babinet-meta}(d),
embedding a complementary structure, as shown in Fig.~\ref{fig:babinet-meta}(b),
and the reflection coefficient is given by $\tilde{r}^{\rm(c)}_i=-2(\tilde{\tau}^{\rm (c)}_i-1/2)$.
Hence, Eqs.~(\ref{Babinet_trans}) and (\ref{Babinet_trans2}) yield the following relations:
\begin{align}
  \tilde{r}\sub{TE} + \tilde{r}\sub{TM}^{\rm (c)} = 0, \label{Babinet_ref}\\
  \tilde{r}\sub{TM} + \tilde{r}\sub{TE}^{\rm (c)} = 0. \label{Babinet_ref2}
\end{align}
These relationships can be regarded as Babinet's relations extended to reflective metasurfaces.
Note that the relations hold under specific conditions for the refractive index $n=2$
and thickness $2d=\pi/k$ for the substrate.
For oblique incidence, 
the condition $2d$ is modified to $2d=\pi/(k\cos\theta)$,
where $\theta$ is the refraction angle in the substrate,
as shown in Appendix \ref{sec:app}.

\begin{figure}[t] 
  \begin{center}
    \includegraphics{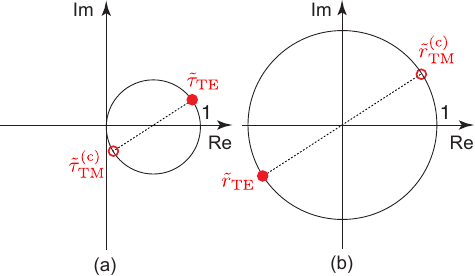}
    \caption{(a) Transmission coefficients of the transmissive metasurface 
    and its complementary one on a complex plane.
    (b) Reflection coefficients of the reflective metasurface 
    and its complementary one on a complex plane.}
    \label{fig:complex}
  \end{center}     
\end{figure}

We reconsider Babinet's relations for transmissive and reflective metasurfaces by representing 
the transmission and reflection coefficients on a complex plane.
For an arbitrary transmissive metasurface composed of planar lossless metals
with a transmission coefficient of $\tilde{\tau}$ and reflection coefficient $\tilde{\rho}$,
continuity of electric fields on the metasurface requires that $1+\tilde{\rho}=\tilde{\tau}$
and energy conservation requires that $|\tilde{\tau}|^2+|\tilde{\rho}|^2=1$.
These two conditions yield 
\begin{align}
   |\tilde{\tau}|^2 + |1-\tilde{\tau}|^2 = 1,
\end{align}
which implies that the transmission coefficient $\tilde{\tau}$ is located on
a circle with a radius of 1/2 and  center of $1/2$ on a complex plane.
Hence, the transmission coefficients $\tilde{\tau}\sub{TE}$ and $\tilde{\tau}\sub{TM}^{\rm (c)}$
for the transmissive metasurfaces as shown in Figs.~\ref{fig:babinet-meta}(a) and (b),
which require the conventional Babinet's relations given by Eqs.~(\ref{Babinet_trans}) and  (\ref{Babinet_trans2}),
are located on the opposite sides of the circle, as shown in Fig.~\ref{fig:complex}(a).
As clearly shown in Fig.~\ref{fig:complex}(a),
the transmissive metasurfaces suffer from reduced transmission 
for large transmission phases ${\rm arg}(\tilde{\tau})$.
This substantially restricts the efficiency of the wave plate operation,
which requires large phase retardation between two polarizations.
The reduction in transmission is attributed to the reflection loss.
By contrast, the reflection coefficient $\tilde{r}$ of an arbitrary reflective metasurface 
composed of planar lossless metals and reflection mirror
satisfies $|\tilde{r}|=1$ because all incident energy is reflected by the metasurfaces
following the interaction between the metallic structures and the mirror.
From Eq.~(\ref{map}),
the transmission coefficients $\tilde{\tau}\sub{TE}$ and $\tilde{\tau}\sub{TM}^{\rm (c)}$
of the transmissive metasurfaces are mapped onto the corresponding reflection coefficients
$\tilde{r}\sub{TE}$ and $\tilde{r}\sub{TM}^{\rm (c)}$, as shown in Fig.~\ref{fig:complex}(b).
It is confirmed that 
Babinet's relation extended to the reflective metasurfaces, which is given 
by Eq.~(\ref{Babinet_ref}), is satisfied.
In contrast to transmissive metasurfaces,
the reflectances of the reflective metasurfaces are always unity
regardless of the phase of reflection,
and polarization manipulation with a high efficiency can be expected.

\section{Half-wave plate using reflective metasurface 
embedding self-complementary structures\label{sec:single}}

\subsection{Babinet's relations for reflective metasurfaces embedding self-complementary structures}

\begin{figure}[b] 
  \begin{center}
    \includegraphics{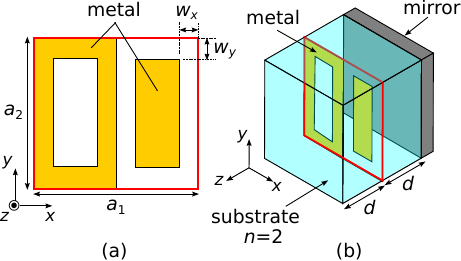}
    \caption{(a) Unit cell of self-complementary structures composed of planar metallic sheet. (b) Unit cell of reflective metasurface embedding
    self-complementary structures.}
    \label{fig:single}
  \end{center}      
\end{figure}

If a structure is congruent with its complementary structure, 
it is referred to as a self-complementary structure.
A typical example of the self-complementary structure, employed in our experiment,
is shown in Fig.~\ref{fig:single}(a).
If the self-complementary structures are embedded in a reflective metasurface
as shown in Fig.~\ref{fig:single}(b),
the extended Babinet's relations given by Eqs.~(\ref{Babinet_ref}) and (\ref{Babinet_ref2})
are reduced to
\begin{align}
  \tilde{r}\sub{TE} + \tilde{r}\sub{TM} = 0, \label{self_comp_babinet}
\end{align}
owing to the self-complementarity $\tilde{r}\sub{TE}=\tilde{r}\sub{TE}^{\rm(c)}$ and
$\tilde{r}\sub{TM}=\tilde{r}\sub{TM}^{\rm(c)}$.
This relationship indicates that  reflective metasurfaces with self-complementary structures
function as half-wave plates,
which exhibit a $\pi$-phase difference between two orthogonal linear polarizations
in a reflection mode.
If the diffraction and dissipation losses are negligible,
the reflectance is expected to be high, $|\tilde{r}\sub{TE}| =| \tilde{r}\sub{TM} | \sim 1$,
which leads to the implementation of efficient half-wave plates.

\subsection{Simulations}

\begin{figure}[b] 
  \begin{center}
    \includegraphics{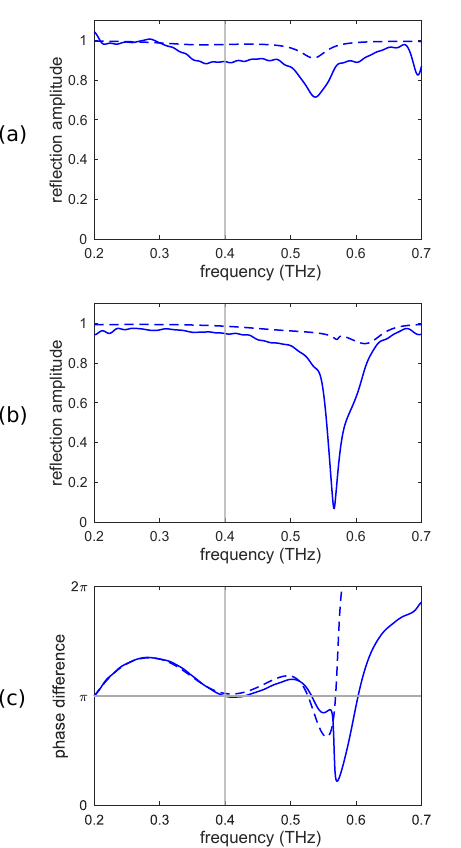}
    \caption{ Amplitude and phase responses of  reflective metasurface with $w_x=30$\mum \,
    and $w_y=50$\mum. Solid lines and dashed lines represent experimental and 
    simulation results, respectively.
    (a) Reflection amplitude spectra $|\tilde{r}\sub{TE}|$ for TE waves. 
    (b) Reflection amplitude spectra $|\tilde{r}\sub{TM}|$ for TM waves.
    (c) Phase difference ${\rm arg}(\tilde{r}\sub{TE}/\tilde{r}\sub{TM})$ between 
    reflected TE waves and TM waves.}
    \label{fig:single-response}
  \end{center}     
\end{figure}

To confirm the extended Babinet's relation given by Eq.~(\ref{self_comp_babinet}),
numerical simulations were first conducted using an electromagnetic simulator (CST MW Studio)
for the metasurface, as shown in Fig.~\ref{fig:single}(b).
The periodicity is assumed to be $a_1=a_2=200$\mum,
and the other dimensions are set to  $w_x=30$\mum, $w_y=50$\mum, and $d=100$\mum.
We assume that the incident plane and incident angle are the $x$-$z$ plane and $\theta\sub{i}=45$\degree,
respectively. 
Under this condition, the lowest diffraction frequency is $c_0/a/(1+\sin\theta\sub{i})=0.88$\,THz.
The incident waves are refracted at the angle of 
$\theta=\sin^{-1} (\sin \theta\sub{i}/n)=20.7$\degree\, in the substrate with the refractive index $n=2$,
and the operating frequency estimated to be $f_0=0.40$\,THz is derived from $2d=\pi/(k\cos \theta)$.
The metal constituting the self-complementary structures was assumed to be aluminum with
a thickness of $300$\,nm, conductivity of $3.56\times10^7$\,S/m,
and the mirror was assumed to be gold with
a thickness of $200$\,nm and conductivity of $4.56\times10^7$\,S/m.
Floquet boundary conditions were applied along the $x$- and $y$-directions
to simulate the interaction between the metasurface and the plane waves incident at 
$\theta\sub{i}=45$\degree.
The input port was located $1000$\mum\, from the top surface of the substrate.

The calculated reflection amplitudes $|\tilde{r}\sub{TE}|$ and $|\tilde{r}\sub{TM}|$ 
for the TE and TM waves are indicated by the dashed lines
in Figs.~\ref{fig:single-response}(a) and (b), respectively.
This frequency region, which is lower than the diffraction frequency $0.88$\,THz,
contains no diffraction; consequently, the reflection amplitudes are close to unity.
The small depressions observed above $0.50$\,THz are caused by resonances 
involving Joule losses in the metallic structures and mirror with finite conductivity.
The dashed line in Fig.~\ref{fig:single-response}(c) shows the calculated phase difference defined by
$\Delta \theta= {\rm arg}(\tilde{r}\sub{TE}/\tilde{r}\sub{TM} )$.
As expected from the extended Babinet's relationship denoted by Eq.~(\ref{self_comp_babinet}),
the phase difference between the TE and TM waves approaches $\pi$ at an operating frequency $0.40$\,THz,
denoted by the gray line.
The slight deviation from $\pi$ is attributed to losses
which are not considered in the derivation of  Babinet's relation.

\subsection{Experimental demonstration}

\begin{figure}[] 
  \begin{center}
    \includegraphics{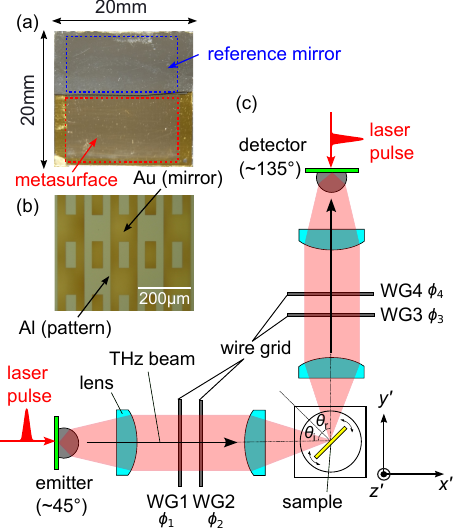}
    \caption{(a) Photograph of the fabricated sample. (b) Microphotograph of the metasurface parts.
    (c) Schematics of experimental setup.}
    \label{fig:setup}
  \end{center}     
\end{figure}

For the experimental demonstration, synthetic quartz was used as the substrate
with a refractive index of $n=2.0$ \cite{Mira2007}.
The metasurface was fabricated using the following procedure.
First, 10nm-thick titanium and 200nm-thick gold films were deposited as a mirror
on a 1mm-thick quartz substrate via electron beam evaporation.
Second, the substrate containing the mirror and a thin quartz substrate were bonded with an 
optical adhesive (Norland, NOA89H).
The thickness of the thin quartz substrate  was $93$\mum,
and the total thickness including the adhesive layer was $108$\mum.
Third, an aluminum film with a thickness of $300$\,nm was deposited via electron beam evaporation.
Self-complementary patterns were fabricated via wet etching
with a mixture comprised of phosphoric acid, nitric acid, acetic acid, and water, at a 16:1:2:1 ratio,
which does not erode the adhesive layer.
Next, the patterned substrate was bonded to a thin quartz substrate,
whose thickness was $92$\mum.
The distance between the embedded aluminum structures and top surface was $104$\mum.
Finally, a $300$nm-thick aluminum film was deposited on half of the sample
to form a reference mirror, which was used to estimate the reflection coefficients. 
A photograph of the fabricated sample is presented in Fig.~\ref{fig:setup}(a).
The reference mirror was placed in the upper half of the sample.
The metasurface embedding the self-complementary structures was placed
in the lower half of that area.
A microphotograph of the metasurface is shown in Fig.~\ref{fig:setup}(b).
The white area corresponds to the aluminum patterns that form self-complementary structures,
and the gold area corresponds to the mirror made of the gold film, 
which can be seen through the aluminum structures.

We employed terahertz time-domain spectroscopy (THz-TDS) 
in the reflective setup as shown in Fig.~\ref{fig:setup}(c)
to acquire the reflection spectra of the fabricated metasurface for TE and TM waves.
The incident terahertz waves propagate along the $x'$ direction,
and the reflected waves from the sample propagate along the $y'$ direction.
In this experiment, the incident angle $\theta\sub{i}$ and reflection angle 
$\theta\sub{r}$ were fixed at $\theta\sub{i}=\theta\sub{r}=45$\degree.
We introduce polarization angles  of $\phi\sub{i}$ and $\phi\sub{r}$
for the incident waves and reflected waves, respectively. 
The polarization directions are defined as $\cos\phi\sub{i} \hat{e}_{y'} + 
\sin \phi\sub{i} \hat{e}_{z'}$ for the incident waves and  $\cos\phi\sub{r} \hat{e}_{x'} + 
\sin \phi\sub{r} \hat{e}_{z'}$ for the reflected waves,
where $\hat{e}_{x'}$, $\hat{e}_{y'}$, and $\hat{e}_{y'}$ are the unit vectors along the
$x'$, $y'$, and $z'$ directions, respectively.
According to this definition, when the incident waves with polarization
$\phi\sub{i}=45$\degree \, are reflected by 
a uniform mirror, the reflected waves have the polarization of $\phi\sub{r}=45$\degree.

The gaps between electrodes of photoconductive antennas used as the emitter and  detector were approximately 
aligned at 45\degree\, and 135\degree, respectively.
The transmission axes of the four wire grids labelled  WG1, WG2, WG3, and WG4 are represented as 
$\phi_1$, $\phi_2$, $\phi_3$, and $\phi_4$, respectively.
We fixed $\phi_1=45$\degree\, and $\phi_4=135$\degree\,
and changed the orientations of WG2 and WG3 depending on the polarizations to be measured.
We set $\phi_2=\phi_3=90$\degree\, for the TE wave measurement and $\phi_2=\phi_3=0$\degree\,
for the TM wave measurement.
The collimated terahertz waves were focused using a lens with a focal length of $50$\,mm,
and the beam diameter of the focused waves at the sample position was $1.7$\,mm,
which was much smaller than the areas of the reference mirror and metasurface.

The sample was mounted on a translation stage along the $z'$ direction.
The reflection signals from the reference mirror $r\sub{ref}(t)$ and 
those from the metasurface $r\sub{meta}(t)$ were acquired by moving the sample vertically. 
The reflection coefficients were obtained as $\tilde{r}(\omega)=-\tilde{r}\sub{meta}(\omega)/\tilde{r}\sub{ref}(\omega)$,
where $\tilde{r}\sub{ref}(\omega)$ and $\tilde{r}\sub{meta}(\omega)$ 
represent the Fourier transforms of $r\sub{ref}(t)$ and $r\sub{meta}(t)$, respectively.
The negative sign was required
because the reference signals $r\sub{ref}(t)$ include the effect of electric-field inversion at the mirror.
The experimental results of the reflection amplitudes for the TE waves $|\tilde{r}\sub{TE}|$ 
and the TM waves $|\tilde{r}\sub{TM}|$ are represented 
by solid lines in Figs.~\ref{fig:single-response}(a) and (b), respectively.
The obtained reflection amplitudes at the operating frequency of $0.40$\,THz were
$|\tilde{r}\sub{TE}|=0.89$ and  $|\tilde{r}\sub{TM}|=0.95$.
The experimental results exhibited resonance features above $0.5$\,THz,
which led to a significant reduction in the reflection.
The dissipation was mainly caused by the dielectric loss of the adhesive layers,
which were not considered in the simulations.
The solid line in Fig.~\ref{fig:single-response}(c) shows 
the phase difference $\Delta \theta= {\rm arg}(\tilde{r}\sub{TE}/\tilde{r}\sub{TM} )$
derived from the experimental results.
The experimental (solid line) and simulation results (dashed line) are in good agreement 
except at the resonant frequencies above $0.5$\,THz.
As expected, the phase difference at the operating frequency $0.40$\,THz approached $\pi$.
The discrepancy around the resonance is mainly attributed to dielectric losses, as discussed above.

\begin{figure}[b] 
  \begin{center}
    \includegraphics{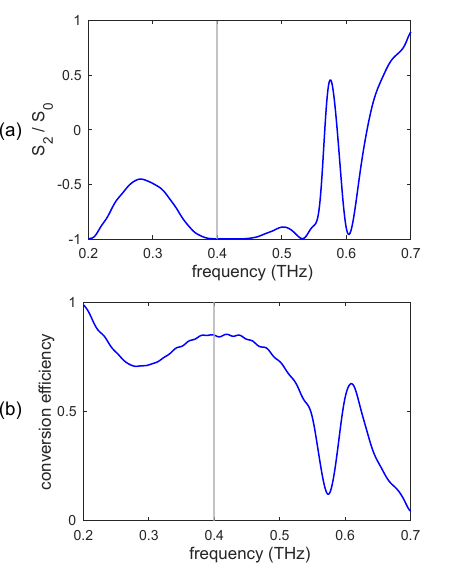}
    \caption{Polarization property of  waves reflected from 
    the reflective metasurface with $w_x=30$\mum \,
    and $w_y=50$\mum\, for  incident waves with $45$\degree\, polarization.
    (a) Normalized Stokes parameter $S_2/S_0$.
    (b) Polarization conversion efficiency from  $\phi\sub{i}=45$\degree\, to $\phi\sub{r}=135$\degree.}
    \label{fig:single-pol}
  \end{center}     
\end{figure}

To estimate the performance of the metasurface as a half-wave plate,
we analyzed the amplitude and polarization of output waves for incident waves 
with $45$\degree\, polarization ($\phi\sub{i}=45$\degree).
Polarization can be characterized by the Stokes parameters \cite{Saleh};
we considered two Stokes parameters, $S_0$ and $S_2$, which can be calculated as
\begin{align}
    S_0 = \frac{|\tilde{r}\sub{TE}|^2+|\tilde{r}\sub{TM}|^2}{2}, \quad 
    S_2 = {\rm Re} (\tilde{r}\sub{TE}^* \tilde{r}\sub{TM}), \label{Stokes}
\end{align}
where $S_0$ denotes the total flux of the output waves
and $S_2$ represents the flux difference between $45$\degree\, polarization and 
$135$\degree\, polarization.
The normalized Stokes parameter defined by $S_2/S_0$ represents the polarization purity,
which becomes $-1$ when the polarization of the output waves is oriented along $135$\degree,
that is $\phi\sub{r}=135$\degree.
Figure~\ref{fig:single-pol}(a) shows the normalized Stokes parameter $S_2/S_0$ calculated using 
experimental results.
At the operating frequency denoted by the gray line, $S_2/S_0$ becomes $-1$ as expected,
which validates that the metasurface functions as a half-wave plate
with an ideal phase difference.

Next, we consider polarization conversion efficiency $\eta$ from $\phi\sub{i}=45$\degree\,
to $\phi\sub{r}=135$\degree.
Reflection coefficients are introduced as 
$\tilde{r}\sub{co}$ for $\phi\sub{r}=45$\degree\, and $\tilde{r}\sub{cr}$ for $\phi\sub{r}=135$\degree,
and the Stokes parameters of the reflected waves are given as 
$S_0=|\tilde{r}\sub{co}|^2+|\tilde{r}\sub{cr}|^2$ and $S_2=|\tilde{r}\sub{co}|^2-|\tilde{r}\sub{cr}|^2$
\cite{Saleh}.
The polarization conversion efficiency $\eta$  can be calculated as
\begin{align}
    \eta \equiv |\tilde{r}\sub{cr}|^2 = \frac{S_0-S_2}{2}.
\end{align}
Figure \ref{fig:single-pol}(b) shows the $\eta$ derived from the experimental results.
The conversion efficiency at the operating frequency, denoted by the gray line, was estimated to be 0.85.
It was confirmed that the pure polarization of $\phi\sub{r}=135$\degree\, 
was achieved with a relatively high conversion efficiency
despite the simple implementation of the metasurface using adhesives to bond the substrates.
A higher conversion efficiency can be realized using an adhesive-free bonding technique such as 
room-temperature bonding between metallic films \cite{Kobachi2021}.

\section{Simultaneous control of wavefront and polarization\label{sec:gradient}}

\subsection{Design of phase-gradient metasurface}

\begin{figure}[h] 
  \begin{center}
    \includegraphics{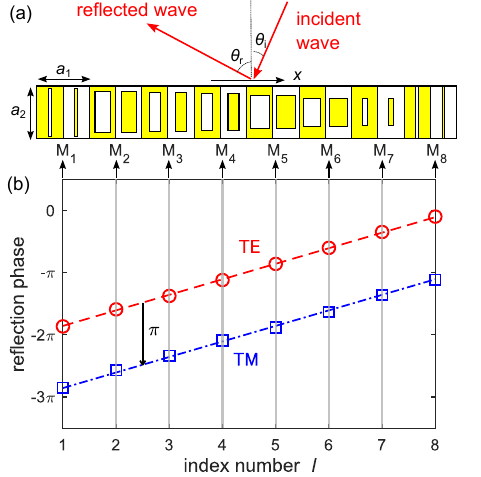}
    \caption{(a) Unit cell (supercell) of the phase-gradient reflective 
    metasurface composed of eight elements ${\rm M}_l$.
    (b) Reflection phases of each element ${\rm M}_l$ for TE and TM waves.}
    \label{fig:gradient}
  \end{center}      
\end{figure}  

The extended Babinet's relation expressed in Eq.~(\ref{self_comp_babinet}) 
holds for reflective metasurfaces embedding arbitrary self-complementary structures.
The phase difference is always $\pi$, however,
either phase ${\arg} (\tilde{r}\sub{TE})$ or ${\arg} (\tilde{r}\sub{TM})$
can be freely selected by designing the self-complementary structures.
If the self-complementary structures depend on the position $x$ and 
${\arg} (\tilde{r}\sub{TE})$ has a constant phase gradient 
$\dd \{{\rm \arg} (\tilde{r}\sub{TE} )\} / \dd x \neq 0$, 
incident TE waves are reflected at anomalous angles that are different from 
the normal reflection angle on a flat mirror \cite{Liu2022}.
The phase gradient of the reflected TM waves is identical to that of the reflected TE waves because of
the extended Babinet's relation Eq.~(\ref{self_comp_babinet}).
Consequently, the TM waves are also reflected at the same anomalous angle.
In the process of the anomalous reflection,
the function as a half-wave plate, that is, ${\rm \arg} (\tilde{r}\sub{TE}) -
{\rm \arg} (\tilde{r}\sub{TM})=\pi$, is maintained.
Hence, the phase-gradient metasurface simultaneously achieves
anomalous reflection and polarization control
by tailoring the phase response for either the TE or TM waves. 
Figure \ref{fig:gradient}(a) shows a unit supercell composed of eight elements labeled as
${\rm M}_l (l=1,2, ..., 8)$, 
with the same design as that shown in Fig.~\ref{fig:single}(a) for various design parameters 
$w_x$ and $w_y$.
If the reflection phase $\psi_l$ for ${\rm M}_l$ is adjusted to satisfy
$\psi_l =  \pi l/4 + \psi'$ ($\psi'$: constant),
which provides a phase difference of $2\pi$ at both ends of the supercell,
the anomalous reflection angle $\theta\sub{r}$ for the incident angle $\theta\sub{i}$
is determined using the following generalized Snell's law:
\begin{align}
  p(\sin \theta\sub{i} - \sin \theta\sub{r})=-\lambda_0, \label{g_snell}
\end{align}
where $p$ is the periodicity of the supercell and
$\lambda_0$ is the wavelength in vacuum \cite{Yu2011}.
For a general periodic structure with  periodicity $p$,
the diffraction angles $\theta\sub{r}$ of $m$-th order are determined by
\begin{align}
p(\sin \theta\sub{i} - \sin \theta\sub{r}) = m \lambda_0. \label{diffraction}
\end{align}
For a phase-gradient metasurface satisfying Eq.~(\ref{g_snell}),
the reflection with a diffraction order $m=-1$ is expected.
For experimental demonstration using the THz-TDS setup shown in Fig.~\ref{fig:setup}(c),
which requires $\theta\sub{i}+\theta\sub{r}=90$\degree,
the parameters were set as follows:
$a_1=a_2=200$\mum, $p=8a_1=1600$\mum, and $d=100$\mum,
which lead to $\theta\sub{i}=24.8$\degree\, and $\theta\sub{r}=65.2$\degree\, for an operating frequency of
$f_0=0.384$\,THz.

\begin{table}
  \caption{
    Design parameters $w_x$ and $w_y$ for each element ${\rm M}_l$
    comprising the phase-gradient reflective metasurface.
    \label{tab:wxwy}}
    \begin{ruledtabular}
  \begin{tabular}{ccccccccc}
     & ${\rm M}_1$  & ${\rm M}_2$ & ${\rm M}_3$ & ${\rm M}_4$
    & ${\rm M}_5$ &${\rm M}_6$ & ${\rm M}_7$ &${\rm M}_8$ \\ \hline
    $w_x$ (\mum) & 44 & 20 & 28 & 28 & 12 &16 & 40 & 44  \\ 
    $w_y$ (\mum) & 8 & 20 & 24 & 28 & 36 & 44 & 48 & 0  \\
   \end{tabular}
  \end{ruledtabular}
  \end{table}

Numerical simulations of the metasurfaces were conducted using the same setup as in Fig.~\ref{fig:single}(b)
for the incident angle $\theta\sub{i}=24.8$\degree,
changing $w_x$ from 0 to 48\mum\, and $w_y$ from 0 to 96\mum\, in increments of 4\mum\,
and singled out eight sets of $w_x$ and $w_y$ from the 325 parameter sets to satisfy the 
linear phase gradient  expressed as $\psi_l =  \pi l/4 + \psi'$. 
The parameter sets are listed in Table \ref{tab:wxwy},
and the calculated reflection phases for the elements ${\rm M}_l$ are denoted 
by circles and squares for the TE and TM waves, respectively, in Fig.~\ref{fig:gradient}(b).
The dashed line was obtained by fitting the reflection phases $\psi_l$ for the TE waves
with $\psi_l = \pi l/4 + \psi'$.
The dashed-dotted line was obtained by shifting the dashed line by $-\pi$.
The reflection phases denoted by the squares are well aligned with the dashed-dotted line,
which implies that each element works as a half-wave plate owing to the self-complementary
structures satisfying Eq.~(\ref{self_comp_babinet}).
Because the phase gradients of the reflected TE and TM waves are identical,
waves incident at $\theta\sub{i}=24.8$\degree\, with arbitrary polarization are reflected anomalously 
at $\theta\sub{i}=65.2$\degree\, while acquiring a $\pi$-phase difference between the TE and TM wave 
components. 

\subsection{Simulations of anomalous reflection and polarization conversion}

\begin{figure}[b] 
  \begin{center}
    \includegraphics{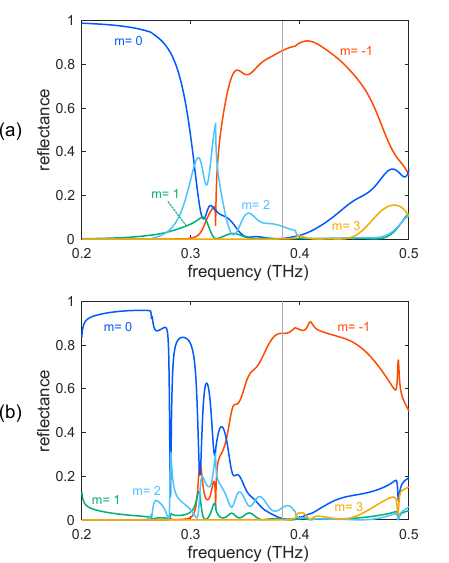}
    \caption{Reflectances to diffraction modes $m=-1, 0, 1, 2, 3$ for (a) TE waves and (b) TM waves. }
    \label{fig:gradient-ref}
  \end{center}     
\end{figure}

Subsequently, we performed numerical simulations 
for the entire metasurface with the supercell as shown in Fig.~\ref{fig:gradient}(a),
and obtained all the reflection coefficients for every diffraction mode.
Figures \ref{fig:gradient-ref}(a) and (b) represent
the reflectances to possible diffraction modes,
which are limited from $m=-1$ to $m=3$ below $0.5$\,THz,
for TE and TM waves, respectively.
The reflectances for $m=-1$ correspond to the designed anomalous reflection,
and reflection angle is $\theta\sub{r}=65.2$\degree\, at 
the operating frequency $f_0=0.384$\,THz denoted by gray lines.
For both polarizations, the reflectances for $m=-1$ were maximized at approximately $0.4$\,THz,
while the reflections of the other modes were significantly suppressed. 
The difference between the designed frequency $f_0=0.384$\,THz\, and the optimal frequency $\sim 0.4$\,THz
can be attributed to interactions between neighboring elements that
were not considered in the design procedure for obtaining the linear phase gradient
in the simulation.
The reflections of the other modes can be further suppressed by 
increasing the number of elements in the supercell
to obtain a smoother phase gradient.

\begin{figure}[b] 
  \begin{center}
    \includegraphics{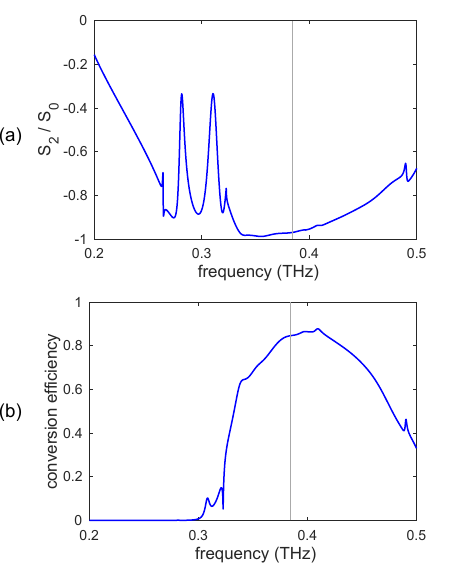}
    \caption{Polarization property of reflected waves in the $m=-1$ mode
    for  incident waves with an illumination angle of $\theta\sub{i}=24.8$\degree\, 
    and  polarization of $\phi\sub{i}=45$\degree.
    (a) Normalized Stokes parameter $S_2/S_0$.
    (b) Polarization conversion efficiency $\eta\sub{s}$ from  $\phi\sub{i}=45$\degree\, to 
    $\phi\sub{r}=135$\degree.}
    \label{fig:gradient-pol}
  \end{center}     
\end{figure}

We can estimate the polarization of the anomalously reflected waves
for the incident waves with $45$\degree\, polarization
using Stokes parameters defined in Eq.~(\ref{Stokes}) 
with the reflection coefficients for the $m=-1$ mode.
Figure \ref{fig:gradient-pol}(a) shows a normalized Stokes parameter $S_2/S_0$ for the $m=-1$ mode.
The value of $S_2/S_0$ is $-0.97$ at the operating frequency of $f_0=0.384$\,THz denoted by the gray line,
which implies that the difference between the reflection phases for the TE and TM modes was close to $\pi$.
Figure \ref{fig:gradient-pol}(b) shows a conversion efficiency $\eta\sub{s}$
from  incident waves with $45$\degree\, polarization to  anomalously reflected waves 
($m=-1$) with $135$\degree\, polarization.
The conversion efficiency $\eta\sub{s}$ was estimated to be 0.85 at the 
operating frequency of $f_0=0.384$\,THz denoted by the gray line.
These simulation results validate that the wavefront control 
could be achieved while maintaining its function as a half-wave plate.

\subsection{Experimental demonstration of anomalous reflection and polarization conversion}

\begin{figure}[t] 
  \begin{center}
    \includegraphics{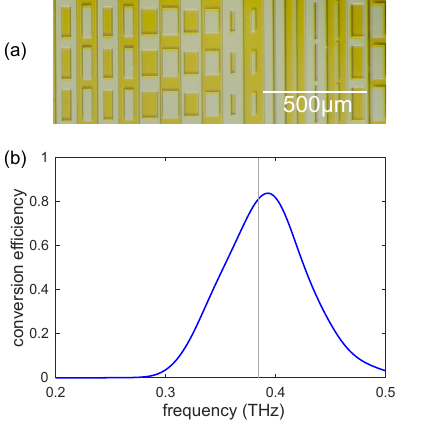}
    \caption{(a) Microphotograph of the fabricated phase-gradient metasurface. 
    (b) Polarization conversion efficiency $\eta\sub{e}$ from the incident waves with
    $\theta\sub{i}=24.8$\degree\, and $\phi\sub{i}=45$\degree\, to 
    anomalously reflected waves with  $\theta\sub{r}=65.2$\degree\, and $\phi\sub{r}=135$\degree.} 
    \label{fig:gradient-exp}  
  \end{center}     
\end{figure}

We fabricated a reflective metasurface using the parameters listed in Table \ref{tab:wxwy}
using the  procedure described in Sec.~\ref{sec:single}.
A microphotograph of the fabricated metasurface is presented in Fig.~\ref{fig:gradient-exp}(a).
We conducted an experiment using the THz-TDS system shown in Fig.~\ref{fig:setup}(c)
to evaluate the polarization conversion efficiency $\eta\sub{e}$
from waves incident at $\theta\sub{i}=24.8$\degree\, with a polarization of $\phi\sub{i}=45$\degree\,
to waves anomalously reflected at $\theta\sub{r}=65.2$\degree\, with a polarization of $\phi\sub{r}=135$\degree.
The angles of wire grids WG1, WG2, WG4 were fixed at $\phi_1=45$\degree, $\phi_2=45$\degree,
and $\phi_4=90$\degree, respectively. 
Under this condition, the polarization of the incident waves was fixed at $\phi\sub{i}=45$\degree.

Prior to measuring the anomalous reflection on the metasurface,
we estimated the reference signal $r\sub{ref}(t)$ using a flat mirror on the upper half of the sample
under the condition where the incident angle $\theta\sub{i}$, 
reflection angle $\theta\sub{r}$, and angle of WG3 $\phi_3$, were $45$\degree. 
The flat mirror reflected the incident waves while maintaining a polarization of $\phi\sub{r}=45$\degree,
which passes through WG3 at $\phi_3=45$\degree.
The polarization of the waves from WG3 was projected along $\phi_4=90$\degree\, by WG4,
and projected at $135$\degree\, by the detector. 
As a result, the amplitude decreased by a factor of two after WG3.
The obtained signal is denoted as $r\sub{ref}(t)$.

Next, the angle of WG3 was changed to $\phi_3=135$\degree.
We rotated the sample holder to satisfy $\theta\sub{i}=24.8$\degree\, and $\theta\sub{r}=65.2$\degree.
The sample holder was moved vertically until the incident waves interacted with the metasurface.
Incident waves with a polarization of $\phi\sub{i}=45$\degree\, interacted 
with the metasurface at an angle of 
$\theta\sub{i}=24.8$\degree\, and were anomalously reflected at an angle of $\theta\sub{r}=65.2$\degree.
A component with a polarization of $135$\degree\, was selected from the anomalously reflected waves
when the waves passing through WG3 aligned along $\phi_3=135$\degree.
The polarization of the waves from WG3 was projected along $\phi_4=90$\degree\, by WG4,
and projected at $135$\degree\, by the detector. 
The amplitude decreased by a factor of two after WG3.
The signal obtained is defined as $r\sub{meta}(t)$.  
From $\tilde{r}\sub{meta}(\omega)$ and $\tilde{r}\sub{ref}(\omega)$,
which are Fourier transforms of $r\sub{meta}(t)$ and $r\sub{ref}(t)$, respectively,
the conversion efficiency can be estimated as $\eta\sub{e}=|\tilde{r}\sub{meta}(\omega)/\tilde{r}\sub{ref}(\omega)|^2$,
which includes the polarization conversion efficiency from $\phi\sub{i}=45$\degree\, to $\phi\sub{r}=135$\degree\,
and diffraction efficiency  from $\theta\sub{i}=24.8$\degree\, to $\theta\sub{r}=65.2$\degree.

The conversion efficiency $\eta\sub{e}$ obtained from the experimental results is shown in Fig.~\ref{fig:gradient-exp}(b).
The value of $\eta\sub{e}$ at the operating frequency of $f_0=0.384$\,THz was estimated to be 0.81.
The optimal frequency was blue-shifted from $f_0=0.384$\,THz. 
This deviation can be attributed to interactions between neighboring elements.
Note that the definitions of conversion efficiencies 
$\eta\sub{s}$ and $\eta\sub{e}$ differ.
While $\eta\sub{s}$ in the simulation
is defined for a specific diffraction order $m=-1$ given by Eq.~(\ref{diffraction}),
thus providing a frequency-dependent  diffraction angle,
$\eta\sub{e}$ in the experiment is defined for a fixed reflection angle of $\theta\sub{r}=65.2$\degree.
The definitions of $\eta\sub{s}$ and $\eta\sub{e}$ coincide only at the operating frequency of $f_0=0.384$\,THz.
Therefore, the shape shown in Fig.~\ref{fig:gradient-pol}(b) differs
from that of Fig.~\ref{fig:gradient-exp}(b).

Compared with previous studies on phase-gradient reflective metasurfaces with similar functions
that realize a $\pi$-phase difference between two linear polarizations
by adjusting the shapes of unit structures \cite{Yang_2014,Ding_2015},
the proposed metasurface always functions as a half-wave plate according to the extended Babinet's relations.
Hence, only one of the TE- or TM-wave reflection phases need to be controlled 
by adjusting the dimension of the self-complementary structures.

Previous research based on conventional Babinet's relations demonstrated
a phase-gradient transmissive metasurface with self-complementary structures,
which separates  incident beams into two non-diffracted beams and two anomalously 
diffracted beams depending on the circular polarizations \cite{Kuznetsov2021}.
In contrast, the phase-gradient refractive metasurface proposed in this paper
reflects most incident waves in an anomalous direction
with an accompanying $\pi$-phase difference between the two linear polarizations.

\section{Conclusion\label{sec:conclusion}}

We  extended the conventional Babinet's relations derived for transmissive metasurfaces
to those of reflective metasurfaces embedding planar metallic structures 
in a substrates with a reflection mirror.
We  proposed two types of reflective metasurfaces embedding self-complementary structures
as  applications of the extended Babinet's relations,
which guarantee $\pi$-phase difference between 
the two orthogonal linear polarizations.
We have theoretically and experimentally demonstrated a metasurface-based half-wave plate
with a reflective metasurface that included single self-complementary structures
in the terahertz region. 
We also designed a reflective metasurface embedding self-complementary structures
with a phase gradient in the reflection,
and demonstrated simultaneous implementation of anomalous reflection and polarization conversion
with high efficiency.
The absorption loss caused mainly by the adhesive layers could be reduced 
by using an adhesive-free bonding technique 
such as the room-temperature bonding between metallic films \cite{Kobachi2021}.
The Joule loss in the metallic structures might be a problem 
especially in higher frequency regions including optical regime. 
In fact, deviations from Babinet's relations have been discussed for transmissive  metasurfaces \cite{Zentgraf_2007, Julian_2021}.
The similar discussion might be applied to the reflective metasurfaces,
because Eq.~(\ref{map}) remains valid even in the presence of the Joule loss.
The reflections of the undesired diffraction modes can be further suppressed by 
increasing the number of elements in the supercell to obtain a smoother phase gradient
or optimizing the entire structure by considering the interactions among the
constituting elements.

In our design of the reflective metasurfaces,
the metallic structures are assumed to be embedded in the substrate,
because the conventional Babinet's relations are rigorously justified for the structures in a uniform medium.
Actually, Babinet's relations for transmissive metasurfaces are empirically valid 
even for metallic structures placed on the top of a substrate.
It might be possible to approximately realize the extended Babinet's relations 
without embedding metallic structures in substrates, but
such implementation is beyond the scope of this paper.

In this paper, we focused on static reflective metasurfaces with self-complementary structures
to verify the extended Babinet's relation given by Eq.~(\ref{self_comp_babinet}),
which is a special case satisfying $\tilde{r}\sub{TE}=\tilde{r}\sub{TE}^{\rm(c)}$ and 
$\tilde{r}\sub{TM}=\tilde{r}\sub{TM}^{\rm(c)}$ in Eqs.~(\ref{Babinet_ref}) and (\ref{Babinet_ref2}).
If the planar structures embedded in the substrate can be converted to their complementary ones,
the general Babinet's relations given by Eqs.~(\ref{Babinet_ref}) and (\ref{Babinet_ref2})
can be demonstrated.
In fact, reconfigurable polarization devices with transmissive metasurfaces,
whose structures can be switched between the original and complementary structures,
have been demonstrated by employing conventional Babinet's relations given by  
Eqs.~(\ref{Babinet_trans}) and (\ref{Babinet_trans2}) \cite{Nakata2019,Nakanishi2020}.
These approaches can be applied to the reflective metasurfaces proposed in this study.
The extended Babinet's relations may pave the way 
for applications in highly efficient multifunctional 
metasurfaces such as reconfigurable intelligent surfaces \cite{Yang2022,elight_Li,PIER_Liu,PIER_Jin}
and polarization-encoded metasurfaces \cite{Lee2019,Ding2021}.
In addition, the relationship expressed by Eq.~(\ref{map}), which is valid for any planar metallic structures
under the derived conditions,
could be utilized for novel design of reflective metasurfaces
by incorporating the design strategy for transmissive metasurfaces.

\begin{acknowledgments}

We thank Y.~Nakata, Y.~Urade, K. Takano, and F. Miyamaru
for helpful discussions and suggestions.
The fabrication of the samples was conducted with the help of Kyoto University Nanotechnology Hub in 
``Advanced Research Infrastructure for Materials and Nanotechnology Project''
sponsored by the Ministry of Education, Culture, Sports, Science and Technology (MEXT), Japan.
This work was supported by JSPS KAKENHI (Grants No.~23K04612).

\end{acknowledgments}

\appendix

\section{Reflection coefficient of reflective metasurface for normal incidence\label{sec:normal}}

\begin{figure}[b] 
  \begin{center}
    \includegraphics{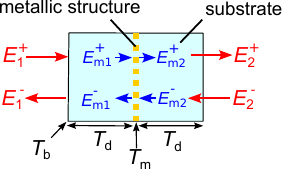}
    \caption{Transfer-matrix elements from the top surface of the substrate to the bottom end of the substrate.}
    \label{fig:normal}
  \end{center}     
\end{figure}

To express $\tilde{r}_{i}$  as a function of $\tilde{\tau}_{i}$ and $\tilde{\rho}_{i}$,
we calculated the transfer matrix $T_i$, defined as 
\begin{align}
  \begin{pmatrix}
    E_2^+ \\ E_2^-
  \end{pmatrix} = T_i
  \begin{pmatrix}
    E_1^+ \\ E_1^-
  \end{pmatrix} = 
  \begin{pmatrix}
    T_{11} & T_{12} \\ T_{21} & T_{22}
  \end{pmatrix} 
  \begin{pmatrix}
    E_1^+ \\ E_1^-
  \end{pmatrix}, 
   \label{Ti}
\end{align}
which connects electric fields $E_1^+$ and  $E_1^-$ at the substrate surface 
with electric fields $E_2^+$ and  $E_2^-$ immediately
before the mirror, as defined in Fig.~\ref{fig:normal}.
The transfer matrix can be decomposed into three elements:
transfer matrix $T\sub{b}$ at the boundary between the vacuum and substrate;
$T\sub{d}$ characterizing the propagation in the substrate of the length $d$;
$T\sub{m}$ on the metallic structures.
$T\sub{b}$ and $T\sub{d}$ are given as
\begin{align}
  & T\sub{b} = \frac{1}{2n}
  \begin{pmatrix}
    n+1 & n-1 \\ n-1 & n+1
  \end{pmatrix}, \\
  & T\sub{d} =
  \begin{pmatrix}
    \ee^{-\jj k d} & 0 \\ 0 & \ee^{\jj k d} 
  \end{pmatrix},
\end{align}
where $n$ is the substrate refractive index 
and $k$ is the wavenumber in the substrate.
The scattered electric fields from the metallic structures satisfy
$E\sub{m2}^+=\tilde{\tau}_i E^+\sub{m1} +\tilde{\rho}_i E^-\sub{m2}$ 
and $E\sub{m1}^-=\tilde{\rho}_i E^+\sub{m1} +\tilde{\tau}_i E^-\sub{m2}$,
where $E\sub{m1}^+$ and $E\sub{m1}^-$ ($E\sub{m2}^-$ and $E\sub{m2}^+$)
represent incoming and outgoing waves on the left (right) side of the structures
respectively, as shown in Fig.~\ref{fig:normal}, and  the transfer matrix $T\sub{m}$
is given as follows \cite{Saleh}:
\begin{align}
  T\sub{m} = \frac{1}{\tilde{\tau}_i}
  \begin{pmatrix}
    \tilde{\tau}_i^2-\tilde{\rho}_i^2 & \tilde{\rho}_i \\ -\tilde{\rho}_i & 1
  \end{pmatrix}. 
   \label{Tm}
\end{align}
If $d$ is assumed to be one-fourth the wavelength in the medium, that is $kd=\pi/2$, then
the whole transfer matrix given by 
$T_i = T\sub{d} T\sub{m} T\sub{d} T\sub{b}$ is expressed as
\begin{align}
  T_i =   \frac{1}{2n\tilde{\tau}_i}
  \begin{pmatrix}
    -(n+3) \tilde{\tau}_i + 2 & -(n-3) \tilde{\tau}_i - 2 \\
    -(n+1) \tilde{\tau}_i + 2 & -(n-1) \tilde{\tau}_i - 2 
  \end{pmatrix}. 
\end{align}
In this calculation, we used $1+\tilde{\rho}_i=\tilde{\tau}_i$,
which is derived from the electric field continuity 
at the metallic structures.
The boundary condition at the mirror surface requires that $E_2^+ + E_2^-=0$.
The reflection coefficient of the metasurface can be derived as
\begin{align}
    \tilde{r}_i = - \frac{T_{11}+T_{21}}{T_{12}+T_{22}} = - \frac{(n+2)\tilde{\tau}_i-2}{(n-2)\tilde{\tau}_i+2}.
    \label{ri}
\end{align}

\section{Proof for the case of oblique incidence\label{sec:app}} 

Under the condition of no scattering to diffraction modes, the electromagnetic property of metasurfaces 
is effectively described by surface impedance or surface admittance \cite{Pfeiffer2014}. 
For the metasurface with uniform structures,
the transmission coefficient for normal incidence is written as
\begin{align}
    \tilde{\tau} = \frac{2\bar{Z}}{2\bar{Z}+Z},
\end{align}
where $Z$ denotes the wave impedance of the surrounding medium
and $\bar{Z}$ is the surface impedance of the metasurface.
If the structures are anisotropic along the $x$ and $y$ directions, as shown
in Figs.~\ref{fig:babinet-meta}(a) or (b),
the surface impedances in the $x$ and $y$ directions can be defined as
$\bar{Z}\sub{TM}$ and  $\bar{Z}\sub{TE}$, respectively.
Additionally, if the incident waves are illuminated at an oblique angle $\theta$,
the effective wave impedances should be modified to $Z/\cos\theta$ and $Z\cos \theta$
for TE  and TM waves, respectively.
As a result, the transmission coefficients for the TE and TM waves are respectively expressed as
\begin{align}
  \tilde{\tau}\sub{TE} = \frac{2 \bar{Z}\sub{TE} \cos \theta}{2 \bar{Z}\sub{TE} \cos \theta + Z},
  \quad \tilde{\tau}\sub{TM} = \frac{2 \bar{Z}\sub{TM}}{2 \bar{Z}\sub{TM}+ Z \cos \theta}.
  \label{tTE}
\end{align}
Similarly, the surface impedances of the complementary structures are introduced as 
$\bar{Z}\sub{TE}^{\rm (c)}$ and  $\bar{Z}\sub{TM}^{\rm (c)}$ for TE and TM waves, respectively,
and the transmission coefficients can be expressed as
\begin{align}
  \tilde{\tau}\sub{TE}^{\rm (c)} = 
  \frac{2 \bar{Z}\sub{TE}^{\rm (c)} \cos \theta}{2 \bar{Z}\sub{TE}^{\rm (c)} \cos \theta + Z},
  \quad \tilde{\tau}\sub{TM}^{\rm (c)} 
  = \frac{2 \bar{Z}\sub{TM}^{\rm (c)}}{2 \bar{Z}\sub{TM}^{\rm (c)}+ Z \cos \theta}.
  \label{tTM}
\end{align}
By substituting these expressions into 
Babinet's relations given by Eqs.~(\ref{Babinet_trans})
and (\ref{Babinet_trans2}),
we obtained the following relationship:
\begin{align}
  \bar{Z}\sub{TE} \bar{Z}\sub{TM}^{\rm (c)} = \frac{Z^2}{4},
  \quad \bar{Z}\sub{TM} \bar{Z}\sub{TE}^{\rm (c)} = \frac{Z^2}{4}. \label{Z_relation}
\end{align}

Next, we derived the reflection coefficients $\tilde{r}_i$ ($i={\rm TE, TM}$) for the reflective metasurface
shown in Figs.~\ref{fig:babinet-meta}(c) and (d) for oblique incidence at the angle of $\theta\sub{i}$.
Assuming that the refractive angle is $\theta$ in the substrate, Snell's law requires 
that $\sin \theta\sub{i}=n\sin \theta$.
The transfer matrices $T\sub{b}$ at the substrate surface are expressed as
\begin{align}
    &T\sub{b} = \frac{1}{2n \cos \theta}
    \begin{pmatrix}
      \cos \theta\sub{i} + n \cos \theta & -\cos \theta\sub{i} + n \cos \theta \\
      -\cos \theta\sub{i} + n \cos \theta & \cos \theta\sub{i} + n \cos \theta
    \end{pmatrix},
  \end{align}
  for TE waves and
  \begin{align}
    &T\sub{b} = \frac{1}{2n \cos \theta}
    \begin{pmatrix}
      \cos \theta + n \cos \theta\sub{i} & -\cos \theta + n \cos \theta\sub{i} \\
      -\cos \theta + n \cos \theta\sub{i} & \cos \theta + n \cos \theta\sub{i}
    \end{pmatrix},
\end{align}
for TM waves \cite{Saleh}.
The transfer matrix for propagation in the substrate is given by
\begin{align}
  T\sub{d} =
  \begin{pmatrix}
    \ee^{-\jj k d \cos \theta} & 0 \\ 0 & \ee^{\jj k d \cos \theta} 
  \end{pmatrix},
\end{align}
for both TE and TM waves.
The transfer matrices for the metallic structure $T\sub{m}$ can be calculated
substituting Eqs.~(\ref{tTE}) and (\ref{tTM}) into Eq.~(\ref{Tm}) with $1+\tilde{\rho}_i=\tilde{\tau}_i$.
Subsequently, the whole transfer matrices $T_i = T\sub{d} T\sub{m} T\sub{d} T\sub{b}$ can be calculated.
Similarly, when Eq.~(\ref{ri}) is derived, the reflection coefficients 
under a condition of $k d \cos \theta=\pi/2$
can be obtained as
\begin{align}
    &\tilde{r}\sub{TE} = \frac{- n^2 \bar{Z}\sub{TE} \cos^2 \theta + Z_0 \cos \theta\sub{i}} 
    {n^2 \bar{Z}\sub{TE} \cos^2 \theta + Z_0 \cos \theta\sub{i}}, \label{rTEo}\\
    &\tilde{r}\sub{TM} = \frac{- n^2 \bar{Z}\sub{TM} \cos \theta\sub{i} + Z_0 \cos^2 \theta}
    {n^2 \bar{Z}\sub{TM} \cos \theta\sub{i} + Z_0 \cos^2 \theta},
\end{align}
for TE and TM waves, respectively.
The reflection coefficients of the refractive metasurface embedding the complementary structures
with the surface impedance $\bar{Z}\sub{TE}^{\rm (c)}$ and  $\bar{Z}\sub{TM}^{\rm (c)}$ are
\begin{align}
  &\tilde{r}\sub{TE}^{\rm (c)} = \frac{- n^2 \bar{Z}\sub{TE}^{\rm (c)} \cos^2 \theta + Z_0 \cos \theta\sub{i}}
  {n^2 \bar{Z}\sub{TE}^{\rm (c)} \cos^2 \theta + Z_0 \cos \theta\sub{i}}, \\
  &\tilde{r}\sub{TM}^{\rm (c)} = \frac{- n^2 \bar{Z}\sub{TM}^{\rm (c)} \cos \theta\sub{i} + Z_0 \cos^2 \theta}
  {n^2 \bar{Z}\sub{TM}^{\rm (c)} \cos \theta\sub{i} + Z_0 \cos^2 \theta}\label{rTMoc}.
\end{align}
From Eqs.~(\ref{rTEo}) and (\ref{rTMoc}), the sum of $\tilde{r}\sub{TE}$ and $\tilde{r}\sub{TM}^{\rm (c)}$ yields
\begin{align}
  &\tilde{r}\sub{TE} + \tilde{r}\sub{TM}^{\rm (c)} =\nonumber \\
  &\frac{2 \cos^2 \theta \cos \theta\sub{i} (Z_0^2 - n^4 \bar{Z}\sub{TE} \bar{Z}\sub{TM}^{\rm (c)})}
  {(n^2 \bar{Z}\sub{TE} \cos^2 \theta + Z_0 \cos \theta\sub{i})(n^2 \bar{Z}\sub{TM}^{\rm (c)} \cos \theta\sub{i} + Z_0 \cos^2 \theta)}.  
  \label{rte_rtm}
\end{align}
Using Eq.~(\ref{Z_relation}), which is derived from Babinet's relations and $Z=Z_0/n$, 
where $Z_0$ is the wave impedance of vacuum, we obtain
\begin{align}
  Z_0^2 - n^4 \bar{Z}\sub{TE} \bar{Z}\sub{TM}^{\rm (c)} = Z_0^2
  \left( 1- \frac{n^2}{4} \right),
\end{align}
and this factor becomes zero for $n=2$,
which yields $\tilde{r}\sub{TE}+\tilde{r}\sub{TM}^{\rm (c)}=0$.
Similarly, $\tilde{r}\sub{TM}+\tilde{r}\sub{TE}^{\rm (c)}=0$ can be verified under this condition.
In summary, the extended Babinet's relations expressed in Eqs.~(\ref{Babinet_ref}) and (\ref{Babinet_ref2})
are satisfied for $n=2$ and $2d=\pi/k\cos\theta$.  

\nocite{*}


%

\end{document}